\documentclass[12pt]{iopart}
\usepackage{iopams}
\usepackage{epsf}
\begin{document}
\title[Dynamics of nonequilibrium quasiparticles]{Dynamics of nonequilibrium quasiparticles in a
double superconducting tunnel junction detector}
\author{M. Ejrnaes,
C. Nappi\footnote[3]{To whom correspondence should be addressed (c.nappi@cib.na.cnr.it)}, and R. Cristiano}
\address{CNR-Istituto di Cibernetica "E. Caianiello", via Campi Flegrei, 34 - Comprensorio "A. Olivetti" - I-80078 Pozzuoli (NA), Italy and \\ 
INFN, Complesso Universitario di Monte Sant'Angelo 
via Cintia, I-80126, Napoli, Italy}
\begin{abstract}
We study a class of superconductive radiation detectors in which the absorption of energy occurs in a long superconductive 
strip while the redout stage is provided by superconductive tunnel junctions positioned at the two ends of the strip. Such a device is capable both of imaging and energy resolution. In the established current scheme, well studied from the theoretical and experimental point of view, a fundamental ingredient is considered the presence of \emph{traps}, 
or regions adjacent to the junctions made of a superconducting material of lower gap. We reconsider the problem by investigating the dynamics of the radiation induced excess quasiparticles in a simpler device, i.e. one without traps.  The  nonequilibrium excess quasiparticles  can be seen to obey a diffusion equation whose coefficients are discontinuous functions of the position. Based on the analytical solution to this  equation, we follow the dynamics of the quasiparticles in the device, predict the signal formation of the detector and discuss the  potentiality offered by this configuration. 
\end{abstract}
\pacs{85.25.Oj, 85.25.Am, 74.40.+k, 74.50.+r}
\maketitle
\section{Introduction}
Superconducting single photon radiation detectors, such as transition edge sensors microbolometers, superconducting tunnel junctions, kinetic 
inductance detectors, and superconductive resonator detectors are attractive devices because they have nearly achieved the desired performance 
for energy-resolved detection of individual photons or particles over a broad energy range with high counting rates \cite{LTD10, book}.

Current work is generally directed towards two objectives: improving energy resolution and scaling to cover large areas. 
For microbolometers the current trend is to try to multiplex an array of detectors to one readout system. 
Various techniques have already demonstrated their viability, although several difficulties remain to be solved.
 The multiplex approach is much more difficult in the case of STJ detectors, as the signals are much faster, 
thus increasing the bandwidth necessary for the multiplexing circuit. A possibility to simplify the problem 
is to drastically reduce the number of channels that has to be readout without reducing the number of pixels
 and the area covered by the array. This can be accomplished by substituting several single pixel STJs 
in the array with long and narrow strips where radiation is absorbed (as long as possible to cover the largest area).
 Each strip is readout through two STJs placed at the ends\cite{Kraus, Jochum, Friedrich}. 
Such a structure is sometime referred in literature with the acronym DROID which stands for 
Distributed Read Out Imaging Device \cite{den}. In DROIDs energy resolution is obtained in 
the same way as for a single STJ by counting the total number of quasiparticles that tunnel 
through the barriers of the two junctions in time coincident pulses; the position is 
obtained by measuring the quasiparticles separately collected  by the two junctions. 
The DROID scheme has been demonstrated to give very good position resolution along the absorber strip, and also a good energy resolution\cite{Li}. 

Since the very beginning when DROIDs were proposed, they were designed by using a higher gap superconductor 
for the absorber and a lower gap superconductor for the STJs electrodes\cite{Kraus}. 
The idea was to exploit the quasiparticle trapping and the multiplication effects\cite{Booth} 
to improve  the collection efficiency. In a DROID quasiparticles produced in the absorber after 
photon absorption will diffuse toward the opposite sides where they are soon trapped  and then \emph{counted}
 by the tunnel junctions. 
There are currently two different ways on how to realize such \emph{traps}
 on the sides of the absorber. One places the trap on top of the absorber \cite{Kraus, den, Nuss, Ville}, 
the other places the trap laterally to the absorber \cite{Friedrich,Lis}. The first method has the advantage 
of the best possible interface between the absorber and the trap material as they can be both deposited \emph{in situ}
 without breaking vacuum during fabrication, but constrains the volume of the trap and junction area severely,
 and results in STJs affected by the proximity effect. The second has almost no limitations on trap design, 
and junction area, the STJ are not affected by the proximity effect, but much care must be taken in 
fabrication to keep the interface between the trap and absorber defect-free.
In both designs energy resolution has been measured as degraded because of the temperature rise 
of the trap when a large amount of quasiparticles enter \cite{Segall, Koz1}. This degrades the 
ability of the trap to efficiently capture the quasiparticles and increases the dark current of 
the STJ, which results into a degraded position and energy resolution. 
The easiest way to remedy this problem is to make bigger traps, if possible,
 which would reduce the heating effect of the quasiparticles. 
Apart from the case where the trap volume is constrained by the 
absorber design, larger traps also means that losses and diffusion 
inside the traps become relevant phenomena that have to be considered.

In the light of the above considerations, we found the physics of a different 
DROID design, one that does not use traps,  interesting to investigate. 
The presence of traps is not a fundamental ingredient that underlies the DROID idea. 
The important issues are the large area coverage with imaging capability and the reduction 
of the readout channels, while maintaining good energy resolution. 
Moreover, single pixel STJ detectors have been made without traps; 
they are well proven examples of detectors that does not have traps but conserve excellent energy resolution\cite{Angl}. 
Large area coverage and reduction of readout channels can be kept by simply 
connecting two STJs to the same bottom electrode (the long absorber) without introducing traps.

A no-trap DROID structure was already analysed in the past \cite{Esposito}, 
but it was not fully appreciated for the advantage of being a simpler structure from 
several points of view: fabrication technology, diffusion, self-heating. 
In the  model contained in \cite{Esposito} the junctions were considered point-like and were only sensitive to the distribution 
function of quasiparticles at the edges of the device
without any disturbance of the nonequilibrium state in the absorber. 
In the present work the analysis has been extended to consider the finite size of the junctions, and to include their influence
on the nonequilibrium dynamics of excess quasiparticles. In this way the presence of the junctions can determine
the detector response.

Finally we remark that the investigation of a DROID structure without traps 
is also of interest in the perspective, just recently emerged\cite{Bramm}, 
that to obtain STJ-detectors which would be really competitive in terms of 
energy resolution one will be probably forced to use for the material of 
the absorber a superconductor with energy gap equal or lower than that of 
aluminium ($\Delta < 0.18$ meV). The reason is that the ultimate energy 
resolution in STJs is proportional to the square root of the energy gap of 
the superconductor where radiation is absorbed. Keeping an energy E = 1 keV 
as reference for X-ray photons, the intrinsic energy resolution requirement 
for advanced applications is 2 eV. This is achieved when the energy gap value 
is equal or less that of the aluminium. Moreover the present state-of-the-art
 in STJ fabrication technology is based on the Al$_2$O$_3$ oxide tunnel barrier, 
indicating again STJs with Al electrodes as the best and possible choice. It is
 then natural to consider a whole aluminium DROID, an Al-absorber laterally readout by 
Al-STJs: definitively a no-trap DROID structure. 

In this paper we describe the response of such a detector under
 the influence of single photons. A model is developed, and an analytic 
expression is obtained for the evolution in time of the one-dimensional 
quasiparticle density, from which we calculate the tunnel currents and 
the collected charge through both the STJs. We discuss the position 
dependence of the pulse height governed by diffusion and quasiparticles decay.

\section{Formulation of the problem}
In a superconductor radiation energy is absorbed in a
complicated process that results in the generation of an excess
number of quasiparticles \cite{Koz2}.
A DROID measures the number of nonequilibrium excess quasiparticles
as  they tunnel through the tunnel junctions positioned at the end
of the superconductive absorber. Each junction is polarised at a convenient voltage point in the subgap
region of the I-V characteristic. When illuminated, the device provides energy and position information of the incoming radiation by comparing the coincident measured charges in the two junctions.
To describe the dynamical behaviour of the proposed
detector a model of the spatial and temporal quasiparticle
excess density is needed.
\begin{figure}[htbp]
\begin{center}
\epsfbox{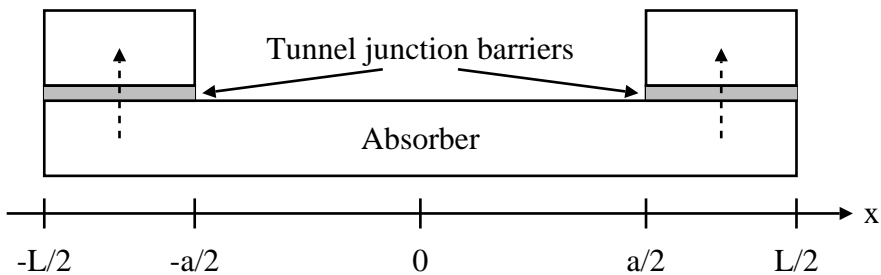}
\end{center}
\caption{\label{fig:geometry} Schematic of the cross section of the proposed detector (not in scale). Dashed arrows on the two sides indicate the excess quasiparticle tunnelling}
\end{figure}
The structure is schematised in figure \ref{fig:geometry}. It consists of
an absorber of length $L$, with  tunnel junctions at the two edges whose length
is $(L-a)/2$. 
We assume that the effect of the tunnel junctions is to remove
the excess quasiparticles created in an impact event of a photon with the absorber with a constant rate probability,
$\gamma_{\rm tunn}$, and that the
quasiparticles do not re-enter the absorber once removed.
We also assume, in the absorber, isotropic diffusion constant, 
$D$, and uniform constant loss rate, $\gamma_{\rm loss}$, which accounts for
 the various mechanism of quasiparticle losses.
These assumptions enable us to describe the  spatial and
temporal evolution of the excess quasiparticle density,
$n\left(x,t\right)$, through the standard diffusion equation \cite{Jochum}\\
\begin{equation}
\label{eqn:diff}
\fl 
\frac{\partial n\left(x,t\right)}{\partial t} =
 D\frac{\partial^2 n\left(x,t\right)}{\partial x^2} -
 \gamma \left( x \right) n\left(x,t\right),
 \quad -{\rm L}/2 <x <{\rm L}/2, \quad 0 < t< \infty
\end{equation}
with the boundary conditions\\
\begin{equation}
\label{eqn:bc}
\frac{\partial n\left( -L/2,t \right)   }{\partial x}
= \frac{\partial n\left(L/2,t \right)   }{\partial x} =0 .
\end{equation}
The presence of the tunnel junctions is then modelled through
$\gamma \left( x \right) = \gamma _{i}$ where the index $i=1,2,3$
refers to the three regions defined as \\
\begin{eqnarray}
\gamma(x) = \left\{ \begin{array}{lll}
 \gamma_{\rm 1}=\gamma_{\rm loss}+\gamma_{\rm tunn} & \textrm{if -L/2 $<$ x $<$ -a/2} &\textrm{Region 1}\\  
 \gamma_{\rm 2}=\gamma_{\rm loss} & \textrm{if -a/2 $<$ x $<$ a/2} &\textrm{Region 2}\\   
 \gamma_{\rm 3}=\gamma_{\rm loss}+\gamma_{\rm tunn} & \textrm{if a/2 $<$ x $<$ L/2} &\textrm{Region 3}
  \end{array} \right.
\end{eqnarray}
Thus in the two junction regions the probability of tunnelling adds
to the probability of loss.
Although simple looking, this problem is more complicated
compared with that arising when traps are present and, as far as we know, the solution has 
never been reported.
We choose to solve this problem by splitting it into three parts,
corresponding to the regions 1, 2, and 3 above.
Thus we need to supplement the boundary conditions, Eq. \ref{eqn:bc},
 with the requirements that
the excess quasiparticle density, and its spatial derivative,
 are continuous functions at $x=\pm a/2$ 
 \begin{eqnarray}
 \label{eqn:bc2}
 \eqalign{
 n_{1}\left(-\frac{a}{2},t\right)=n_{2}\left(-\frac{a}{2},t\right),
 \,n_{2}\left(\frac{a}{2},t\right)=n_{3}\left(\frac{a}{2},t\right)\\
 \frac{\partial n_{1}\left(-\frac{a}{2},t\right)}{\partial x}=\frac{\partial n_{2}\left(-\frac{a}{2},t\right)}{\partial x},\,
 \frac{\partial n_{2}\left(\frac{a}{2},t\right)}{\partial x}=\frac{\partial n_{3}\left(\frac{a}{2},t\right)}{\partial x}}
 \end{eqnarray}

The radiation impact on the absorber is modeled through the
initial condition:
\begin{equation}
\label{eqn:initialcondition}
n \left(x,0  \right)=N_{\rm 0}\delta \left(x-x_{0}  \right) ,
\end{equation}
where $x_{\rm 0}$ is the radiation impact point and $N_0$ the number of quasiparticles generated by 
the radiation impact.
All this provides seven relations which together with equation
(\ref{eqn:diff}) allows to determine the solution in the three
domains between $-L/2$ and $L/2$.

\section{Solution of the problem}
\label{sec:solution}
The solution to the problem, equations (\ref{eqn:diff})-(\ref{eqn:initialcondition}), can be obtained by  the separation
of variables method \cite{Williams}
introducing the arbitrary constant $\nu$. 
Then the unknown density can be written as an integral on all possible values of
$\nu$ in the three previously defined spatial domains.
\begin{eqnarray}
\label{eqn:fullsolution}
n_i\left(x,t\right) & = & \int_{-\gamma_i}^0
  \rme^{-\left(\gamma_i + \nu_i\right)t} \left[X_i\left(\nu_i\right)
  \rme^{\sqrt{\frac{-\nu_i}{D}}x}+Y_i\left(\nu_i\right)
  \rme^{-\sqrt{\frac{-\nu_i}{D}}x} \right]\rmd\nu_i+    \nonumber \\
& & +\int_0^\infty \rme^{-\left(\gamma_i+\nu_i\right)t} \left[
  X_i\left(\nu_i\right) \cos{\sqrt{\frac{\nu_i}{D}}x}+
  Y_i\left(\nu_i\right) \sin{\sqrt{\frac{\nu_i}{D}}x}
  \right]\rmd\nu_i .
\end{eqnarray}

Inserting the full formal solution, equation (\ref{eqn:fullsolution}), into the six boundary conditions, see \ref{app:matrix}, 
we find that the possible values of $\nu$ is an infinite discrete set.
It is convenient to distinguish the sign of $\nu$ by introducing
$\alpha^2=-D\nu/L^2$
and $\mu^2=D\nu/L^2$.
Then the two equations determining these values of $\nu$, written in terms of $\alpha$ and $\mu$   respectively, are
\begin{eqnarray}
\label{eqn:alphavals}
  \left[\gamma_{\rm D} +\left(\gamma_{\rm D}-2\alpha^2\right)
  \cosh\left(\alpha-a\alpha\right)\right]
  \sin\left(a\sqrt{\gamma_{\rm D}-\alpha^2}\right) + &  \nonumber\\
  -2\alpha\sqrt{\gamma_{\rm D}-\alpha^2}
  \cos\left(a\sqrt{\gamma_{\rm D}-\alpha^2}\right)
  \sinh\left(\alpha-a\alpha\right) = 0
\end{eqnarray}
and
\begin{eqnarray}
\label{eqn:muvals}
  \left[\gamma_{\rm D} +\left(\gamma_{\rm D}+2\mu^2\right)
  \cos\left(\mu-a\mu\right)\right]
  \sin\left(a\sqrt{\gamma_{\rm D}+\mu^2}\right) + &     \nonumber\\
  +2\mu\sqrt{\gamma_{\rm D}+\mu^2}
  \cos\left(a\sqrt{\gamma_{\rm D}+\mu^2}\right)
  \sin\left(\mu-a\mu\right) = 0,
\end{eqnarray}
where
$\gamma_{\rm D}=\gamma_{\rm tunn} L^2/D$.
Once the possible values of the constant $\nu$ have been determined  the two integrals in equation (\ref{eqn:fullsolution}) change into sums.
Thus equation (\ref{eqn:fullsolution}) can be written as
\begin{eqnarray}
\label{eqn:fullsolsum}
n_i\left(x,\tau\right)&=&\sum_{n=0}^{N_{\rm \alpha}-1}
  R_n\left(\alpha_n,x_0\right) \rme^{-\left(\frac{\gamma_i}{\gamma_{\rm loss}} -
        \alpha_n^2\frac{\Lambda^2}{L^2}\right)\tau}
  F_{i,n}\left(\alpha_n ,x\right) + \nonumber \\
  & & + \sum_{n=N_\alpha}^\infty
  S_n\left(\mu_n,x_0\right)\rme^{-\left(\frac{\gamma_i}{\gamma_{\rm loss}} +
        \mu_n^2\frac{\Lambda^2}{L^2}\right)\tau}
  G_{i,n}\left(\mu_n ,x\right)
\end{eqnarray}
where $\alpha_n$, $\mu_n$ are solutions to equations (\ref{eqn:alphavals}) and (\ref{eqn:muvals}), respectively, and the two proportionality constants $R_n\left(\alpha_n,x_0\right)$ and
$S_n\left(\mu_n,x_0\right)$ will be determined by the initial
condition. We have also introduced the normalised time,
$\tau=t\gamma_{\rm loss}$, the finite number of possible $\alpha$-values,
$N_\alpha$ (see \ref{app:matrix}), the diffusion length $\Lambda=\sqrt{D/\gamma_{\rm loss}}$,
and the spatial base functions
\begin{eqnarray}
\label{eqn:spatialfuncs}
\fl F_{i,n}\left(\alpha_n ,x\right) = \left\{
  \begin{array}{l}
    A_{i,n}\left(\alpha_n\right) \rme^{\alpha_n x} +
    B_{i,n}\left(\alpha_n\right) \rme^{-\alpha_n x} \\
    A_{i,n}\left(\alpha_n\right) \cos\left(\sqrt{\gamma_D-\alpha_n^2}x\right) +
    B_{i,n}\left(\alpha_n\right) \sin\left(\sqrt{\gamma_D-\alpha_n^2}x\right)
  \end{array}
  \right. \!\!\!\!
  \begin{array}{l}
    , \, i=1,3 \\
    , \, i=2
  \end{array} \\
\fl G_{i,n}\left(\mu_n ,x\right) = \left\{
  \begin{array}{l}
    C_{i,n}\left(\mu_n\right) \cos\left(\mu_n x\right) +
    D_{i,n}\left(\mu_n\right) \sin\left(\mu_n x\right) \\
    C_{i,n}\left(\mu_n\right) \cos\left(\sqrt{\gamma_D+\mu_n^2}x\right) +
    D_{i,n}\left(\mu_n\right) \sin\left(\sqrt{\gamma_D+\mu_n^2}x\right)
  \end{array}
  \right. \!\!\!\!
  \begin{array}{l}
    , \, i=1,3 \\
    , \, i=2
  \end{array}
\end{eqnarray}
in which all lengths ($x,x_0,a$) are normalised to $L$, unless otherwise indicated.
The procedure to obtain the coefficients $A_{i,n},B_{i,n},C_{i,n},D_{i,n},R_n,S_n$ is described in detail in \ref{app:solcoeff}
along with their analytical expressions.

\begin{figure}[htbp]
\begin{center}
\epsfbox{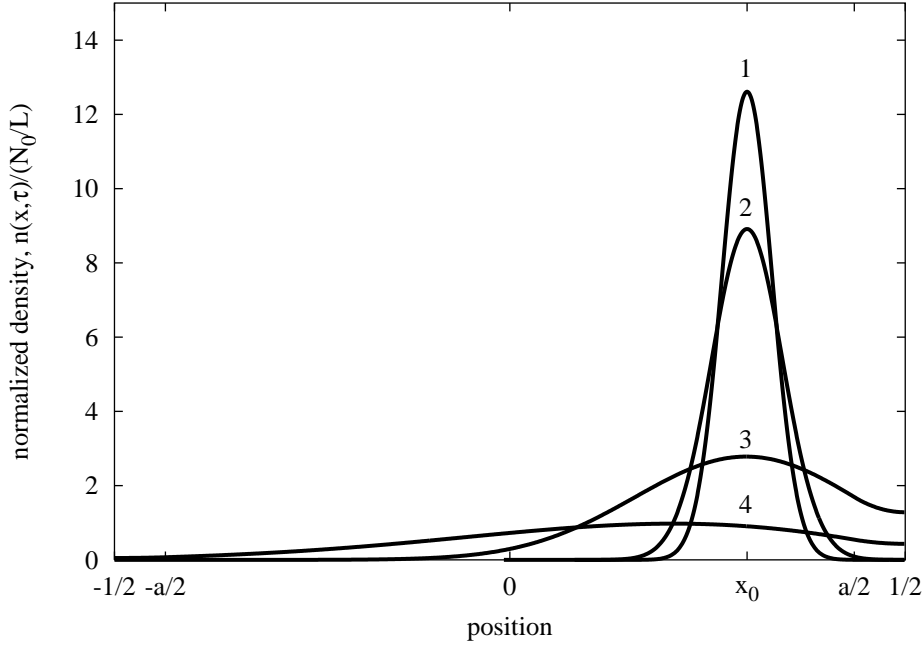}
\end{center}
\caption{\label{fig:tempevol}Temporal evolution of the density of
excess quasiparticles in the device for an event occurring at $x_0=0.3$. Curves $1,2,3,4$ correspond to $\tau =t \gamma_{\rm loss}=0.0005,0.001,0.01,0.5$. The parameters used are $\Lambda/L=1$, $\gamma_{\rm tunn}/\gamma_{\rm loss}=143$,
$a=0.87$}
\end{figure}

At this point the mathematical problem is completely solved, and the
spatial and temporal response of the proposed detector is known.
An example of the spatial evolution of the density of excess
quasiparticles in an aluminium device is shown in figure \ref{fig:tempevol} at
subsequent instants of time (in units of $\gamma_{\rm loss}$) for an event occurred at $x_0=0.3$. 
We used values of the parameters which are compatible with those of a device based on aluminium:
$D=60~\rm cm^2/s$,
$\gamma_{\rm loss}=10^4~s^{-1}$,
  $\Lambda=775~\mu$m,
$L=775~\mu$m,
for the absorber, and  a tunnelling rate
 $\gamma_{\rm tunn}=1.43\times10^6~s^{-1}$
 for the two junctions, each long 
 $50~\mu$m,
 ($a=675~\mu$m, in absolute units). Note that no trace remains, in the time evolving density, of the discontinuity 
 in the coefficient $\gamma(x)$ so that the transition between the absorber and the junction regions is completely smooth.  
 
\section{Analysis of the detector signal}
\label{sec:analys}
\begin{figure}[htbp]
\begin{center}
  \epsfbox{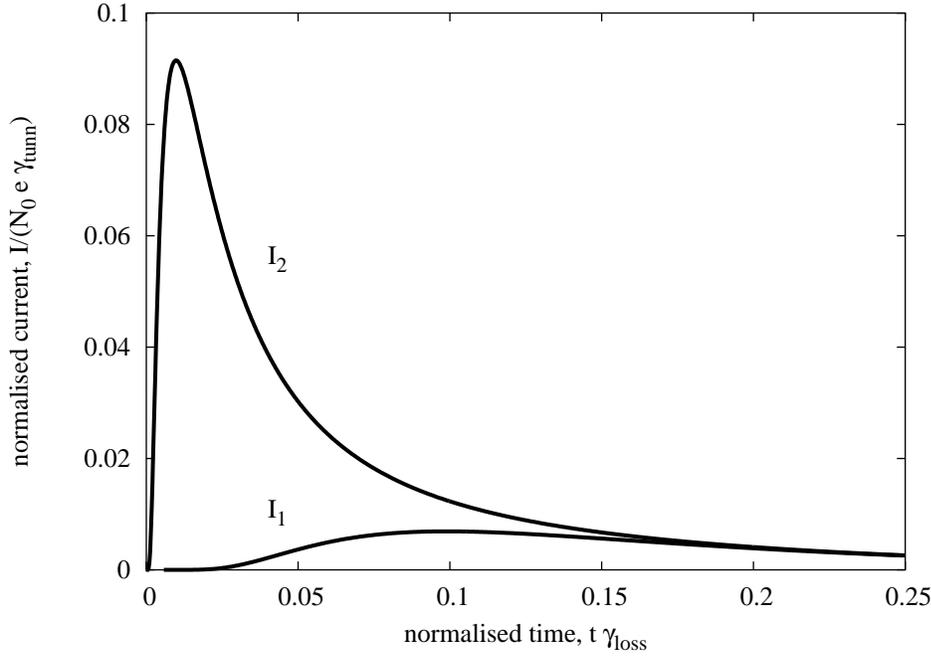}
\end{center}
\caption{\label{fig:i1i2}Temporal evolution of the normalised currents passing through
the tunnel junctions for the absorption event located at $x_0=0.3$. Parameters  correspond to $\Lambda/L=1$, $\gamma_{\rm tunn}/\gamma_{\rm loss}=143$,
$a=0.87$}
\end{figure}

To show the potential of the result obtained in the previous section, we will calculate the signals that are
typically measured in experiments with DROIDs (i.e. tunnelling current pulses flowing through the junctions).
These currents are calculated  as 
\begin{equation}
  \label{eq:i1}
  I_1\left(x_0, \tau\right)=\gamma_{\rm tunn}eL\int_{-\frac{1}{2}}^{-\frac{a}{2}}n\left(x, \tau\right){\rm d}x
\end{equation}
\begin{equation}
  \label{eq:i2}
  I_2\left(x_0, \tau\right)=\gamma_{\rm tunn}eL\int_{\frac{a}{2}}^{\frac{1}{2}}n\left(x, \tau\right){\rm d}x
\end{equation}
and are given by the expressions \\
\begin{eqnarray}
\label{eqn:i1}
\fl  I_1\left(x_0,\tau\right)=\gamma_{\rm tunn}eL
  \sum_{n=0}^{N_{\rm \alpha}-1}
  \frac{R_n\left(\alpha_n,x_0\right)}{\alpha_n}
  \rme^{-\left(\frac{\gamma_{\rm 1}}{\gamma_{\rm loss}} - \alpha_n^2\frac{\Lambda^2}{L^2}\right)\tau}
  \left(\rme^\frac{\alpha_n}{2} - \rme^\frac{a \alpha_n}{2}\right)
  \left(A_{1,n}\rme^{-\alpha_n\frac{1+a}{2}}+B_{1,n}\right)
  \nonumber \\
  + \gamma_{\rm tunn}eL\sum_{n=N_\alpha}^\infty
  \frac{S_n\left(\mu_n,x_0\right)}{\mu_n}
  \rme^{-\left(\frac{\gamma_{\rm 1}}{\gamma_{\rm loss}} + \mu_n^2\frac{\Lambda^2}{L^2}\right)\tau} \times \nonumber \\
  \times \left\{D_{1,n}\left[\cos\left(\frac{\mu_n}{2}\right)-\cos\left(\frac{a\mu_n}{2}\right)\right] +
         C_{1,n}\left[\sin\left(\frac{\mu_n}{2}\right)-\sin\left(\frac{a\mu_n}{2}\right)\right]
  \right\}
\end{eqnarray}
\begin{eqnarray}
\label{eqn:i2}
\fl  I_2\left(x_0,\tau\right)=\gamma_{\rm tunn}eL
\sum_{n=0}^{N_{\rm \alpha}-1}
\frac{R_n\left(\alpha_n,x_0\right)}{\alpha_n}
\rme^{-\left(\frac{\gamma_{\rm 1}}{\gamma_{\rm loss}} - \alpha_n^2\frac{\Lambda^2}{L^2}\right)\tau}
\left(\rme^\frac{\alpha_n}{2} - \rme^\frac{a \alpha_n}{2}\right)
\left(B_{3,n}\rme^{-\alpha_n\frac{1+a}{2}}+A_{3,n}\right) \nonumber \\
+ \gamma_{\rm tunn}eL\sum_{n=N_\alpha}^\infty
\frac{S_n\left(\mu_n,x_0\right)}{\mu_n}
\rme^{-\left(\frac{\gamma_{\rm 1}}{\gamma_{\rm loss}} + \mu_n^2\frac{\Lambda^2}{L^2}\right)\tau} \times \nonumber \\
\times \left\{D_{3,n}\left[\cos\left(\frac{a\mu_n}{2}\right)-\cos\left(\frac{\mu_n}{2}\right)\right]+ C_{3,n}\left[\sin\left(\frac{\mu_n}{2}\right)-\sin\left(\frac{a\mu_n}{2}\right)\right]
\right\}
\end{eqnarray}
Figure \ref{fig:i1i2} shows the two current pulses for the absorption event shown in figure \ref{fig:tempevol}.
In literature one typically finds expressions for the collected charges at the two lateral sides of the absorber, which were directly derived through equations governing the charge itself \cite{Jochum} rather than the density.
We emphasise that to fully characterise the DROID behaviour it is certainly of great help to also have expressions for the extra tunnelling currents as those calculated above, which is possible only when the solution for $n(x,t)$ has been obtained as we did here.

Finally the collected charge in each tunnel junction as a function of the impact point $x_0$ can be
calculated as
\begin{equation}
  \label{eq:charge12}
  Q_{1,2}\left(x_0\right)=\int_0^\infty I_{1,2}\left(x_0, t\right){\rm d}t.
\end{equation}
and is given by the expressions:
\begin{eqnarray}
\label{eqn:q1}
\fl  Q_1\left(x_0\right)=\frac{\gamma_{\rm tunn}}{\gamma_{\rm loss}}eL
  \sum_{n=0}^{N_{\rm \alpha}-1}
  \frac{R_n\left(\alpha_n,x_0\right)}{\alpha_n\left(\frac{\gamma_{\rm 1}}{\gamma_{\rm loss}} -
  \alpha_n^2\frac{\Lambda^2}{L^2}\right)}
  \left(\rme^\frac{\alpha_n}{2} - \rme^\frac{a \alpha_n}{2}\right)
  \left(A_{1,n}\rme^{-\alpha_n\frac{1+a}{2}}+B_{1,n}\right)
  \nonumber \\
  + \frac{\gamma_{\rm tunn}}{\gamma_{\rm loss}}eL\sum_{n=N_\alpha}^\infty
  \frac{S_n\left(\mu_n,x_0\right)}{\mu_n\left(\frac{\gamma_{\rm 1}}{\gamma_{\rm loss}} +
  \mu_n^2\frac{\Lambda^2}{L^2}\right)}
  \left\{D_{1,n}\left[\cos\left(\frac{\mu_n}{2}\right)-\cos\left(\frac{a\mu_n}{2}\right)\right]\right. +
  \nonumber \\
  \left. + C_{1,n}\left[\sin\left(\frac{\mu_n}{2}\right)-\sin\left(\frac{a\mu_n}{2}\right)\right]
  \right\}
\end{eqnarray}
\begin{eqnarray}
\label{eqn:q2}
\fl  Q_2\left(x_0\right)=\frac{\gamma_{\rm tunn}}{\gamma_{\rm loss}}eL
  \sum_{n=0}^{N_{\rm \alpha}-1}
  \frac{R_n\left(\alpha_n,x_0\right)}{\alpha_n\left(\frac{\gamma_{\rm 1}}{\gamma_{\rm loss}} -
  \alpha_n^2\frac{\Lambda^2}{L^2}\right)}
  \left(\rme^\frac{\alpha_n}{2} - \rme^\frac{a \alpha_n}{2}\right)
  \left(B_{3,n}\rme^{-\alpha_n\frac{1+a}{2}}+A_{3,n}\right)
  \nonumber \\
  + \frac{\gamma_{\rm tunn}}{\gamma_{\rm loss}}eL\sum_{n=N_\alpha}^\infty
  \frac{S_n\left(\mu_n,x_0\right)}{\mu_n\left(\frac{\gamma_{\rm 1}}{\gamma_{\rm loss}} +
  \mu_n^2\frac{\Lambda^2}{L^2}\right)}
  \left\{D_{3,n}\left[\cos\left(\frac{a\mu_n}{2}\right)-\cos\left(\frac{\mu_n}{2}\right)\right]\right. +
  \nonumber \\
  \left. + C_{3,n}\left[\sin\left(\frac{\mu_n}{2}\right)-\sin\left(\frac{a\mu_n}{2}\right)\right]
  \right\}
\end{eqnarray}
Through equations (\ref{eqn:q1}),(\ref{eqn:q2}) curves at constant energy and constant absorption positions can be constructed (see next section)  for a direct comparison with the typical experimental data representation  \cite{Kraus,Jochum}.

\section{ Results and discussion}
\begin{figure}[htbp]
\begin{center}
  \epsfbox{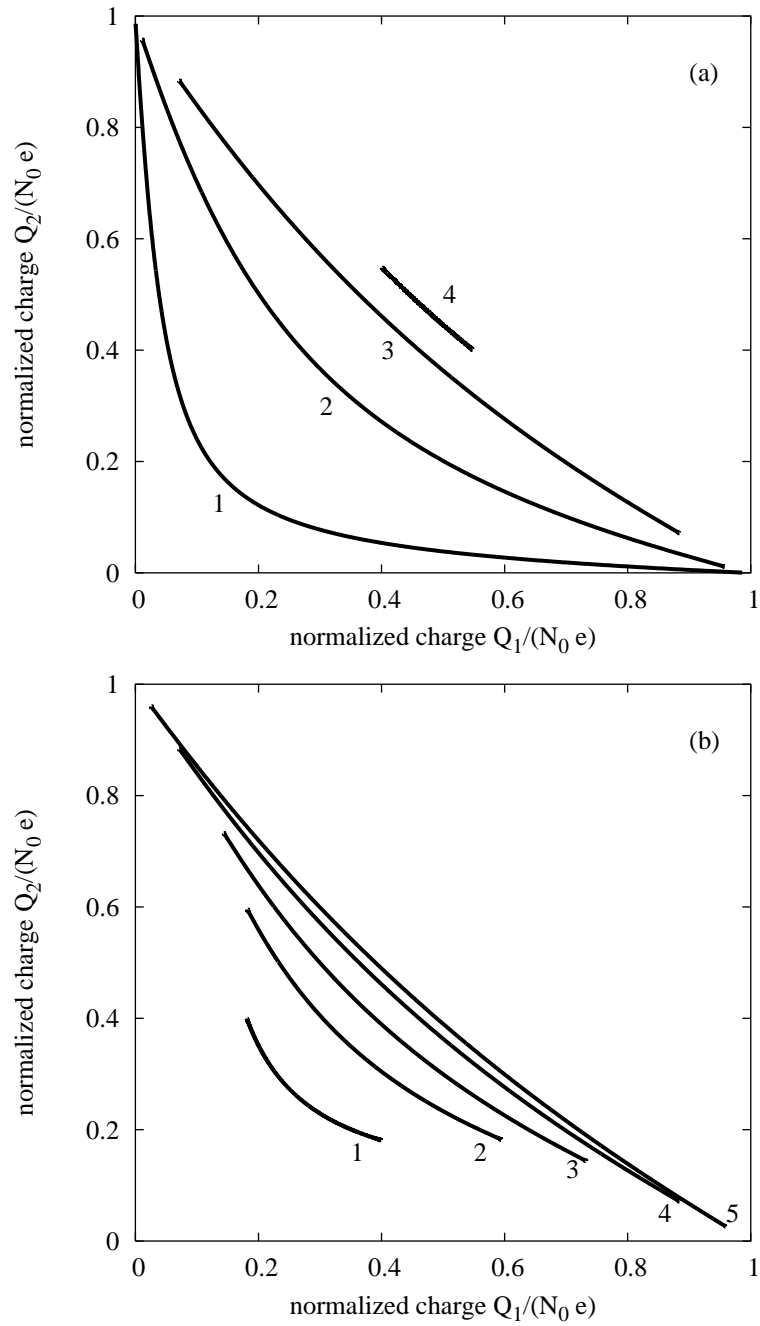}
\end{center}
\caption{\label{fig:q1q2}$\left. a\right)$ Collected charges $Q_2$ vs $Q_1$ for increasing values of $\Lambda/L, (0.25,0.5,1.0,5.0)$, $\gamma_{\rm tunn}/\gamma_{\rm loss}=143$ and $a=0.871$.$\left.b\right)$ Collected charges $Q_2$ vs $Q_1$ for increasing values of $\gamma_{\rm tunn}/\gamma_{\rm loss}$, $(10.0, 25.0, 50.0, 143.0, 400.0)$, $\Lambda/L=1$,$a=0.871$}
\end{figure}
The behaviour of a DROID is often analysed by plotting  coincident collected charges ($Q_1,Q_2$) for all the 
possible values of $x_0$ at constant energy (i.e. constant $N_0$). 
Following this choice, in figure \ref{fig:q1q2}a 
we show the response of the no-trap-DROID (equations (\ref{eqn:q1}) and (\ref{eqn:q2})) for various values of $\Lambda/L$ corresponding 
to absorbers with different values of the diffusion constant $D$ or different values of the loss rate $\gamma_{\rm loss}$.
 The initial and the final points of each curve, where charge maximises, correspond always to the $-L/2$ and $L/2$ positions
 on the absorber. Therefore spatial resolution scales inversly with charge collected in these points. It is seen that increasing $\Lambda /L$, simultaneously  increases the charge yield, makes the charge yield 
more uniform (the curve shape becomes more straight)  and reduces the spatial resolution (the total extension of the curve shortens). 
 In the limit of very efficient diffusion (curve 4), the spatial resolution is lost and the response becomes similar to that of a single STJ. 
Figure \ref{fig:q1q2}b shows the no-trap-DROID response for increasing values of $\gamma_{\rm tunn}/\gamma_{\rm loss}$, corresponding to devices 
with increasing tunnel junction transparencies. 
In this case the charge yield, the uniformity of the charge yield and spatial resolution of the device simultaneously increase.
 Curve 3 in figure \ref{fig:q1q2}a and curve 4 in 
figure \ref{fig:q1q2}b coincide and refer to the same device 
considered in figure \ref{fig:tempevol} and \ref{fig:i1i2}. 
We stress that these curves  predict a behaviour comparable 
to that of a DROID based on the quasiparticle trapping principle 
\cite{Jochum,Friedrich} and that all the essential features of this last device are conserved in the device described here.

\begin{figure}[htbp]
\begin{center}
  \epsfbox{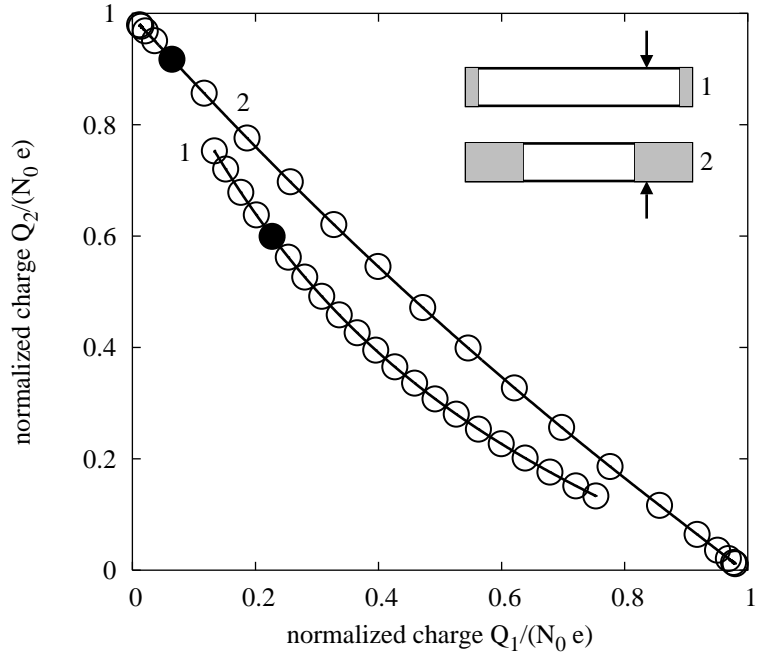}
\end{center}
\caption{\label{fig:q1q2a} Detector response for two different values of the geometrical parameter $a$ (the size of the junction-free region of the device). Curve 1, a device with $a=0.95$, curve 2 a device with $a=0.5$. In the inset, a sketch of the two detectors corresponding to the two curves. The gray regions indicate the junctions. Open circles indicate  signals coming from equi-spaced and uniformly distributed positions along the device. The two solid circles are signals coming from the position marked by the arrows on the two devices}
\end{figure}

It is also of interest to briefly discuss the role of the junction size by investigating the role of the geometrical parameter $a$. 
This can be done by using the same type of  $Q_1, Q_2$ plot as  before by simply emphasising  points corresponding to the same 
equidistant positions of radiation absorption. Two such curves are shown in figure \ref{fig:q1q2a} in which the open circles, 
superimposed to the solid line,  mark the points corresponding to an arbitrary segmentation of the device 
in twenty equal intervals. The lower curve, curve 1, has $a=0.95$, 
the higher, curve 2,  has $a=0.5$, moreover $\gamma_{\rm tunn}/\gamma_{\rm loss}=143$, $\Lambda/L=1$; 
the geometry of the devices corresponding to these curves are sketched in the inset of the same figure 
(the gray regions indicate the junctions).  A smaller $a$ value (larger junctions) results in an increased charge 
yield with larger uniformity. Furthermore the spatial resolution is strongly nonuniform as indicated by the 
 lateral crowding,  and the simultaneous central rarefaction of points. Therefore the DROID becomes very position sensitive
in the central region while under the junctions this sensitivity is degraded. 
To better illustrate this effects, in figure \ref{fig:q1q2a}, we  indicated 
two points (solid circle) on each curve corresponding to the same impinging radiation position (arbitrarily chosen) 
on the two detectors and marked in the inset by arrows.

In conclusion we have analysed in detail the behaviour of a class of superconductive radiation detectors, denominated DROID, in which 
the absorption of energy occurs in a long superconductive strip and the readout stage is provided by superconducting tunnel junctions 
positioned at the two  ends of the strip. We have introduced boundary conditions suitable for the absence of trapping regions 
and solved analytically the resulting diffusion problem. We underline that the model provides not only the total charges 
collected by the two junctions but also explicit expressions for the tunnelling current pulses. Our analysis shows that a device
 based on a single superconductive material (i.e. without traps) is still capable both of position and energy resolution. Recently, 
aluminium has been proposed as the material of choice to achieve the best energy resolution and imaging is obtained by arrays 
of single superconducting tunnel junctions. Unless other considerations are put forward, our work demonstrates the feasibility 
of an alternative viable solution to imaging, i.e the DROID configuration, which can be entirely based on a single material, e.g. aluminium.

\section{Acknowledgements}
This work has been partially supported by the EC-RTM Network \emph{Applied Cryodetectors} contract no. HPRNCT2002000322 
and Progetto FIRB n. RBNE01KJHT.

\clearpage
\appendix 
\section{Matrix coefficients} 
\label{app:matrix} 
Inserting the full formal solution, equation (\ref{eqn:fullsolution}), into
the six boundary conditions, equation (\ref{eqn:bc}), and using the substitutions
$\nu_{\rm 1}=\nu_{\rm 3}=\nu$ and
$\nu_{\rm 2}=\nu+\gamma_{\rm tunn}$, it is found
that time-independence of the boundary conditions splits the set of equations for $X_1,Y_1,X_2,Y_2,X_3,Y_3$ into three sets, 
depending on the value of
$\nu$.
The first set of equations, for the values
$-\gamma_{\rm tunn}-\gamma_{\rm loss}<\nu<-\gamma_{\rm tunn}$, arises
from matching exponential functions in region 1 and 3 with
exponential functions in region 2.
The second set of equations, for the values
$-\gamma_{\rm tunn}<\nu<0$, arises from matching exponential
functions in region 1 and 3 with the trigonometric functions in
region 2.
The last set of equations, for positive values of $\nu$, arises from
matching trigonometric functions in region 1 and 3 with trigonometric
functions in region 2.
Despite the differences between the three sets of equations it is
possible to write them in the same matrix
form 

\begin{equation}
\label{eqn:bcmatrixform}
\left[
 \begin{array}{cccccc}
  m_{11} & m_{12} & 0 & 0 & 0 & 0 \\
  m_{21} & m_{22} & m_{23} & m_{24} & 0 & 0 \\
  m_{31} & m_{32} & m_{33} & m_{34} & 0 & 0 \\
  0 & 0 & m_{43} & m_{44} & m_{45} & m_{46} \\
  0 & 0 & m_{53} & m_{54} & m_{55} & m_{56} \\
  0 & 0 & 0 & 0 & m_{65} & m_{66} \\
 \end{array}
\right]
\times
\left[
  \begin{array}{c}
    X_1 \\ Y_1 \\ X_2 \\ Y_2 \\ X_3 \\ Y_3
  \end{array}
\right]
=
\left[
\begin{array}{c}
 0 \\ 0 \\ 0 \\ 0 \\ 0 \\ 0
\end{array}
\right]
\end{equation}
The exact expression  of the matrix elements $m_{i,j}$ depends on the three possible ranges of values of $\nu$.
They are listed in Table (\ref{tab:melements}), written in terms of $\alpha^2=-D\nu/L^2$ and $\mu^2=D\nu/L^2$.
\begin{table}[b]
\caption{\label{tab:melements} Expressions in terms of $\alpha$ and $\mu$ of the matrix elements \\ appearing in eq. (\ref{eqn:bcmatrixform}).
  ($\alpha^2=-D\nu/L^2$
and $\mu^2=D\nu/L^2$).}
\begin{tabular}{@{}llll}
\br Element&$-\gamma-\gamma_t<\nu<-\gamma_t$&$-\gamma_t<\nu<0$&$\nu>0$\\
\mr
$m_{11}$&$\exp\left(-\frac{\alpha}{2}\right)$&$\exp\left(-\frac{\alpha}{2}\right)$&$\sin\left(\frac{\mu}{2}\right)$\\
$m_{12}$&$-\exp\left(\frac{\alpha}{2}\right)$&$-\exp\left(\frac{\alpha}{2}\right)$&$\cos\left(\frac{\mu}{2}\right)$\\
$m_{21}$&$\exp\left(-\frac{a\alpha}{2}\right)$&$\exp\left(-\frac{a\alpha}{2}\right)$&$\cos\left(\frac{a\mu}{2}\right)$\\
$m_{22}$&$\exp\left(\frac{a\alpha}{2}\right)$&$\exp\left(\frac{a\alpha}{2}\right)$&$-\sin\left(\frac{a\mu}{2}\right)$\\
$m_{23}$&$-\exp\left(-\frac{a\sqrt{\alpha^2-\gamma_D}}{2}\right)$&$-\cos\left(\frac{a\sqrt{\gamma_D-\alpha^2}}{2}\right)$&$-\cos\left(\frac{a\sqrt{\gamma_D+\mu^2}}{2}\right)$\\
$m_{24}$&$-\exp\left(\frac{a\sqrt{\alpha^2-\gamma_D}}{2}\right)$&$\sin\left(\frac{a\sqrt{\gamma_D-\alpha^2}}{2}\right)$&$\sin\left(\frac{a\sqrt{\gamma_D+\mu^2}}{2}\right)$\\
$m_{31}$&$\alpha m_{21}$&$\alpha m_{21}$&$-\mu m_{22}$\\
$m_{32}$&$-\alpha m_{22}$&$-\alpha m_{22}$&$\mu m_{21}$\\
$m_{33}$&$\sqrt{\alpha^2-\gamma_D}m_{23}$&$-\sqrt{\gamma_D-\alpha^2}m_{24}$&$-\sqrt{\gamma_D+\mu^2}m_{24}$\\
$m_{34}$&$-\sqrt{\alpha^2-\gamma_D}m_{24}$&$\sqrt{\gamma_D-\alpha^2}m_{23}$&$\sqrt{\gamma_D+\mu^2}m_{23}$\\
$m_{43}$&$\sqrt{\alpha^2-\gamma_D}m_{24}$&$\sqrt{\gamma_D-\alpha^2}m_{24}$&$\sqrt{\gamma_D+\mu^2}m_{24}$\\
$m_{44}$&$-\sqrt{\alpha^2-\gamma_D}m_{23}$&$\sqrt{\gamma_D-\alpha^2}m_{23}$&$\sqrt{\gamma_D+\mu^2}m_{23}$\\
$m_{45}$&$\alpha m_{22}$&$\alpha m_{22}$&$\mu m_{22}$\\
$m_{46}$&$-\alpha m_{21}$&$-\alpha m_{21}$&$\mu m_{21}$\\
$m_{53}$&$m_{24}$&$m_{23}$&$m_{23}$\\
$m_{54}$&$m_{23}$&$-m_{24}$&$-m_{24}$\\
$m_{55}$&$m_{22}$&$m_{22}$&$m_{21}$\\
$m_{56}$&$m_{21}$&$m_{21}$&$-m_{22}$\\
$m_{65}$&$-m_{12}$&$-m_{12}$&$-m_{11}$\\
$m_{66}$&$-m_{11}$&$-m_{11}$&$m_{12}$\\
\br
\end{tabular}
\end{table}
Inspection of the determinants of the three matrices shows that  for
$-\gamma_{\rm tunn}-\gamma_{\rm loss}<\nu<-\gamma_{\rm tunn}$ only the trivial solution exists. 
On the contrary, non-trivial solutions can be found for a finite 
number, $N_\alpha$, of values of $\nu$ when the interval  $-\gamma_{\rm tunn}<\nu<0$ is considered (solutions to equation (\ref{eqn:alphavals})), and also
for an infinite  discrete set of  values of $\nu$ for $\nu>0$ (solutions to equation (\ref{eqn:muvals})). These two equations express the conditions 
for theexitancee of non-trivial solutions in the two considered intervals of $\nu$.

\clearpage
\section{Coefficients in the solution, equation (\ref{eqn:fullsolsum})}
\label{app:solcoeff}
As the problem is symmetric in space the spatial modes separate in
even and odd modes.
The first mode is even, the second odd and so on.
This means that the symmetry of the first mode in the second sum in
equation (\ref{eqn:fullsolsum}) depends on whether $N_\alpha$ is even
or odd (which is why we have chosen to let is start from $N_\alpha$
instead of from 0).
Insight into when a mode leaves the second sum in equation
(\ref{eqn:fullsolsum}) and appears in the first sum is obtained by
examining equation \ref{eqn:muvals} in the limit
$\mu\rightarrow 0$.
It is found that equation \ref{eqn:muvals} reduces to
$\sin\left(a\sqrt{\gamma_{\rm D}}\right)=0$.
From this limit it is easily seen that it is possible to express
$N_\alpha$ as the smallest integer greater than
${a\sqrt{\gamma_{\rm D}}}/{\pi}.$

We obtain exact expressions for the even coefficients ($A_{i,2n}$ and
$B_{i,2n}$) by assuming $A_{2,2n}$ known.
Then we create a reduced system of five equations with five unknowns
by moving all the terms containing $A_{2,n}$ to the right hand side
in equation (\ref{eqn:bcmatrixform}) and eliminate a row (exactly
which one is not important as the six rows arelinearlyy dependent on
each other).
By solving this system of equations we obtain $A_{i,2n}$ and
$B_{i,2b}$ except for a proportional constant,
$R_n\left(\alpha_n,x_0\right)$ in equation (\ref{eqn:fullsolsum}),
which will be determined using the initial condition (equation
(\ref{eqn:initialcondition})).
Care should be taken as the expressions obtained are not valid for
the odd coefficients ($A_{i,2n+1}$ and $B_{i,2n+1}$) because for
these coefficients $A_{2,2n+1}$ is zero and thus creates problems
when selected as known.
We instead select $B_{2,2n+1}$ as known and repeat the procedure as
above.
Likewise for the second sum in equation (\ref{eqn:fullsolsum}) we
select first $C_{2,2n}$ as known to obtain the even coefficients and
then select $D_{2,2n+1}$ as known to obtain the odd coefficients.
Also for the these coefficients we have to determine a proportional
constant, $S_n\left(\mu_n,x_0\right)$ in equation
(\ref{eqn:fullsolsum}), using equation (\ref{eqn:initialcondition}).
The expressions obtained are:
\begin{eqnarray}
A_{1,n}\left(\alpha_n\right)&=&\exp\left(\alpha_n+\frac{a\alpha_n}{2}\right)\sqrt{\gamma_D-\alpha_n^2}\\
B_{1,n}\left(\alpha_n\right)&=&\exp\left(\frac{a\alpha_n}{2}\right)\sqrt{\gamma_D-\alpha_n^2}\\
C_{1,n}\left(\mu_n\right)&=&\sqrt{\gamma_D+\mu_n^2}\cos\left(\frac{\mu_n}{2}\right)\\
D_{1,n}\left(\mu_n\right)&=&-\sqrt{\gamma_D+\mu_n^2}\sin\left(\frac{\mu_n}{2}\right)\\
A_{2,n}\left(\alpha_n\right)&=&\left[\exp\left(\alpha_n\right)+\exp\left(a\alpha_n\right)\right]\sqrt{\gamma_D-\alpha_n^2}
  \cos\left(\frac{a\sqrt{\gamma_D-\alpha_n^2}}{2}\right) + \\ \nonumber
  & &+ \left[\exp\left(\alpha_n\right)-\exp\left(a\alpha_n\right)\right]\alpha_n
  \sin\left(\frac{a\sqrt{\gamma_D-\alpha_n^2}}{2}\right)\\
B_{2,n}\left(\alpha_n\right)&=&\left[\exp\left(\alpha_n\right)-\exp\left(a\alpha_n\right)\right]\alpha_n
  \cos\left(\frac{a\sqrt{\gamma_D-\alpha_n^2}}{2}\right) + \\ \nonumber
  & &- \left[\exp\left(\alpha_n\right)+\exp\left(a\alpha_n\right)\right]\sqrt{\gamma_D-\alpha_n^2}
  \sin\left(\frac{a\sqrt{\gamma_D-\alpha_n^2}}{2}\right)\\
C_{2,n}\left(\mu_n\right)&=&\sqrt{\gamma_D+\mu_n^2}\cos\left(\frac{\mu_n-a\mu_n}{2}\right)
  \cos\left(\frac{a\sqrt{\gamma_D+\mu_n^2}}{2}\right) + \\ \nonumber
  & & -\mu_n\sin\left(\frac{\mu_n-a\mu_n}{2}\right)\sin\left(\frac{a\sqrt{\gamma_D+\mu_n^2}}{2}\right)\\
D_{2,n}\left(\mu_n\right)&=&-\mu_n\sin\left(\frac{\mu_n-a\mu_n}{2}\right)\cos\left(\frac{a\sqrt{\gamma_D+\mu_n^2}}{2}\right)+\\ \nonumber
  & & -\sqrt{\gamma_D+\mu_n^2}\cos\left(\frac{\mu_n-a\mu_n}{2}\right)
  \sin\left(\frac{a\sqrt{\gamma_D+\mu_n^2}}{2}\right)\\
A_{3,n}\left(\alpha_n\right)&=&\left(-1\right)^n\exp\left(\frac{a\alpha_n}{2}\right)\sqrt{\gamma_D-\alpha_n^2}\\
B_{3,n}\left(\alpha_n\right)&=&\left(-1\right)^n\exp\left(\alpha_n+\frac{a\alpha_n}{2}\right)\sqrt{\gamma_D-\alpha_n^2}\\
C_{3,n}\left(\mu_n\right)&=&\left(-1\right)^n\sqrt{\gamma_D+\mu_n^2}\cos\left(\frac{\mu_n}{2}\right)\\
D_{3,n}\left(\mu_n\right)&=&\left(-1\right)^n\sqrt{\gamma_D+\mu_n^2}\sin\left(\frac{\mu_n}{2}\right) .
\end{eqnarray}
Note that if the second sum in equation \ref{eqn:fullsolsum} is
started at zero instead of $N_\alpha$ then the coefficients
$C_{3,n}$ and $D_{3,n}$ should be multiplied with
$\left(-1\right)^{N_\alpha}$.

The initial condition, equation (\ref{eqn:initialcondition}), can
be used to find $R_n\left(\alpha_n,x_0\right)$ and
$S_n\left(\mu_n,x_0\right)$ as:
\begin{equation}
\label{eqn:initialconst}
  \label{eq:initialconst}
  R_n\left(\alpha_n,x_0\right) =
   N_0\frac{F_{i,n}\left(\alpha_n,x_0\right)}
        {\Gamma_\alpha\left(\alpha_n\right)^2}
  , \, S_n\left(\mu_n,x_0\right) =
    N_0\frac{G_{i,n}\left(\mu_n,x_0\right)}
        {\Gamma_\mu\left(\mu_n\right)^2}
\end{equation}
Where the norm of the modes,
\begin{eqnarray}
\Gamma_\alpha\left(\alpha_n\right)^2 = L
\frac{\rme^{\alpha_n\left(1+a\right)}}{\alpha_n} \left[ \left(2-a\right)\gamma_{\rm
D}\alpha_n -
  2{\alpha_n}^3 +
  a\gamma_{\rm D}\alpha_n\cosh\left(\alpha_n-\alpha_n a\right) + \right.\nonumber \\
  \left. +2\gamma_{\rm D}\sinh\left(\alpha_n-\alpha_n a\right)\right]
\end{eqnarray}
and
\begin{equation}
\fl \Gamma_\mu\left(\mu_n\right)^2 = \frac{L}{4\mu_n}
\left[\left(2-a\right)\gamma_D\mu_n +
  2{\mu_n}^3 +
  a\gamma_D\mu_n\cos\left(\mu_n-a\mu_n\right) +
  2\gamma_D\sin\left(\mu_n-a\mu_n\right)\right]
\end{equation}
have been calculated using the equations
(\ref{eqn:alphavals}) and (\ref{eqn:muvals}) in order to
simplify as much as possible their form.
To obtain equation (\ref{eqn:initialconst}) we have used the
orthogonality of the modes when integrated over the entire length.
The exact expression of equation (\ref{eq:initialconst}) depends on
whether $x_0$ is located in region 1, 2, or 3.

\end{document}